# A Comprehensive Test System for Transmission Expansion Planning: Modeling N−1 Contingencies and Multi-Loading Scenarios

Bhuban Dhamala, *Graduate Student Member, IEEE,* and Mona Ghassemi, *Senior Member, IEEE*

***Abstract*—This paper presents a high-voltage test system designed specifically for transmission expansion planning (TEP) and explores multiple TEP studies using this test system. The network incorporates long transmission lines, lines are accurately modeled, and line parameters are calculated using the equivalent $\pi$ circuit model for long transmission lines to account for the distributed nature of line parameters. The paper provides detailed load flow analyses for both normal and all contingency conditions for three different loading conditions (peak load, dominant load, and light load), demonstrating that the proposed test system offers technically feasible load flow solutions at these loading scenarios. As the real power system is subject to various loading scenarios and should be effectively operable under all conditions, this test system accurately replicates the properties of real power systems. Furthermore, this paper presents multiple TEP cases to supply the load at a new location. TEP cases are conducted with different numbers of transmission line connections, and each case is underscored by its respective maximum capacity satisfying all technical requirements for normal and all single contingencies under three different scenarios. The cost of TEP for each case is calculated and compared in terms of the average cost per MW of power delivered to the new bus.**

## I. Introduction

BASED on the 2023 International Energy Outlook report, global energy consumption is projected to surge by significant margins across various sectors by 2050 [1]. Specifically, the industrial sector is expected to witness a 62% increase, while the transportation sector could see a rise of around 41%. Remarkably commercial and industrial infrastructure may experience a threefold surge in consumption during this period. Additionally, global electricity generation is forecasted to surge by 30% to 76% compared to 2022 levels, predominantly driven by the adoption of zero-emission technologies. These advancements are further fueled by Renewable Portfolio Standards (RPS), aimed at curbing carbon emissions [2]. Utilities and power providers are obligated by these regulations to obtain a designated fraction of their electricity from renewable sources. Amidst the dynamic evolution of power systems, ensuring the uninterrupted and reliable transmission of adequate energy throughout the power network is paramount. With the escalating demand for energy and the increasing integration of extensive renewable sources into the primary grid, establishing a resilient, reliable, and economically viable power grid capable of satisfying both present and future demands becomes ever more pressing. A report suggested that achieving zero emission in the U.S. by 2050 necessitates a tripling of the capacity of high-voltage transmission lines compared to levels it had in 2023 [3]. Meeting such ambitious targets requires substantial investment and extensive studies on transmission expansion planning (TEP). TEP involves making strategic decisions over the long term to reinforce the transmission network by integrating new lines and substations [4]. Effective transmission expansion planning analysis relies heavily on the presence of a base test system. Such a system serves as a fundamental model for analysis and assessment, facilitating the exploration of planning aspects, formulating TEP scenarios, optimizing them using different optimization techniques, and evaluating TEP strategies considering various technical, economical, and environmental factors for practical implementation [5, 6].

Commonly employed in power systems and research, the IEEE test cases [7] hold a venerable position within the field. However, despite their widespread use, these IEEE test cases often inadequately capture the complexities of contemporary power systems for transmission expansion planning owing to

The authors are with the Department of Electrical and Computer Engineering, The University of Texas at Dallas, Richardson, TX 75080 USA (e-mail: mona.ghassemi@ utdallas.edu).

their low voltage levels, short transmission line lengths, and limited loading scenarios. In response to these limitations, many synthetic power grid test cases, designed to emulate real power systems, have been made accessible as steady-state test cases and dynamic test cases. Reference [8] provides the base case details, [9] includes economic data, and [10] contains dynamic information for large synthetic grids covering regions such as Texas, the northeastern United States, the Western United States, and a combined synthetic east and west US grid. These resources are accessible at [11]. These systems range from extra high voltage transmission levels to low voltage distribution levels and include data for load flow analysis, transient analysis, and geomagnetic disturbance analysis. Additionally, reference [12] provides the Texas combined electric gas test case, while a synthetic test case specifically designed for Geomagnetic Disturbance Studies on the Tennessee footprint is presented in reference [13]. There are many other publicly available test systems [14-16]. While these test systems offer features more aligned with real power systems in size and complexity, they predominantly cater to single loading conditions, lacking specificity regarding various load conditions or other potential contingencies. Additionally, uncertainties persist regarding the capacity of these existing test systems to operate effectively under all single contingency conditions. Moreover, these real or synthetic test systems do not provide comprehensive information about the length of transmission lines in the power network and their parameters. Distances between buses or the lengths of lines within these test systems are crucial to accurately estimating the distances between buses intended for integration. Achieving accurate replication of real-world power system conditions and facilitating successful transmission expansion planning is dependent on these factors.

A multitude of test systems have been proposed in the literature for TEP studies across different voltage levels. Among these, the 6-bus Garver system [17], IEEE 24-bus system [18], HRP-38 bus system tailored for TEP studies with high renewable energy integration [19] have been extensively utilized by the researcher in TEP purposes. Furthermore, systems such as the 46-bus Southern Brazilian network [20, 21], the 87-bus Brazilian north-eastern network [22, 23], the Columbian power system comprising 93 nodes [24], along with the IEEE 118 bus test system, also known as the NREL-118 bus system [25], and a reduced version of the Western Electricity Coordinating Council (WECC) system with 300 nodes [26] have gained prominence and are widely used by researchers on different aspects of transmission planning. A large-scale test system specifically for ACTEP has also been proposed in reference [4]. However, most of them are designed for DC TEP studies, and some systems that are proposed for AC TEP do not cover the high voltage level, such as 500 kV. Moreover, there are some test systems specifically tailored for assessing the reliability of transmission and distribution systems, as documented in references [27-31]. These mentioned test systems, whether for TEP studies or reliability evaluations, are limited to one loading scenario and do not guarantee successful system operation by meeting all technical requirements under all single contingencies. Furthermore, their voltage levels are limited to the distribution voltage level or lower voltage transmission levels. Detailed line information, such as line length and line parameters, is not clearly communicated. The absence of clear communication about the capabilities of these test systems poses a significant challenge for transmission expansion planning studies. Achieving both technically and economically optimal transmission expansion planning studies is impossible without ensuring technical feasibility under diverse loading scenarios and under all single contingencies.

This paper addresses the technical gaps by introducing a fictitious test system specifically tailored for TEP studies at a transmission voltage level of 500 kV. The test system includes all high-voltage long transmission lines, carefully crafted to emulate real-world high-voltage power transmission systems, and lines are modeled accurately using an equivalent $\pi$ model to account for the distributed nature of line constraints for long transmission lines. Moreover, multiple TEP cases using the test system are carried out to supply power to the load at a new location, featuring varying numbers of transmission line connections from the nearest possible buses. The objective of the TEP is to ascertain the maximum power deliverable by each case and to calculate the cost for each TEP scenario. TEP cases are then compared in terms of average cost per MW power transfer to the new bus, facilitating the selection of the optimal number of line connections. Both the test system and the system after TEP are engineered to operate effectively under normal condition and all single contingencies, accommodating different loading scenarios – peak load, dominant load, and light load – while adhering to bus voltage, reactive power generation, and line loading criteria.

The structure of this paper is outlined as follows: Section II provides an overview of the test system and details on the modeling of its transmission lines. In Section III, the formulation of load flow analysis is discussed. Section IV offers a detailed examination of power flow analysis within the test system. Section V introduces multiple case studies of TEP. Finally, Section VI concludes the paper by summarizing the findings and insights of the study.

## II. Overview of the Test System

### *A. Network Topology*

Fig. 1 displays the single-line diagram of the proposed test system, comprising 17 buses operating at 500 kV.

Bus 1 is designated as the slack bus, while buses 3, 6, 8, 10, 12, 13, and 15 serve as voltage-controlled buses. The remaining nine buses are designated as load buses. Generators and loads in the system are randomly determined, each approximately 300-400 km apart. Generation stations are spaced at least 300 km apart, with some exceeding 600 km. Every bus is considered to be located in its geographic location, and the length of each transmission line is measured and presented relative to a reference length of 300 km between buses 7 and 12. Detailed information regarding line length and the number of line connections between buses is provided in Table I.

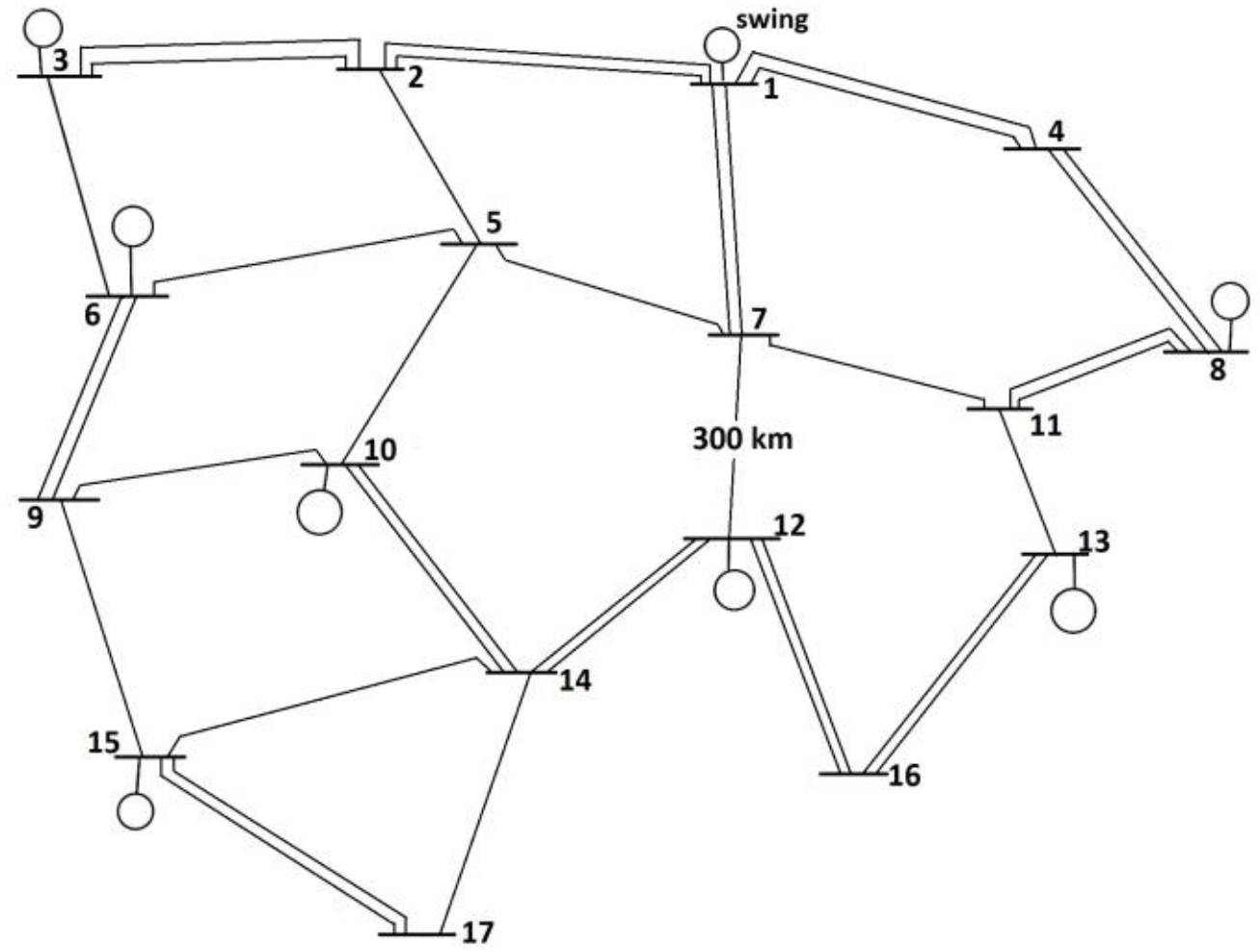

Fig. 1. Single-line diagram of the test system.

TABLE I
TRANSMISSION LINE LENGTHS AND CONNECTED BUSES

| Line (Bus-Bus) | No. of circuits | Length (km) | Line (Bus-Bus) | No. of circuits | Length (km) |
|---|---|---|---|---|---|
| 1-2 | 2 | 410.33 | 7-12 | 1 | 300.05 |
| 1-4 | 2 | 426.78 | 8-11 | 2 | 349.10 |
| 1-7 | 2 | 370.92 | 9-10 | 1 | 447.28 |
| 2-3 | 2 | 436.90 | 9-15 | 1 | 398.20 |
| 2-5 | 1 | 294.56 | 10-14 | 2 | 392.74 |
| 3-6 | 1 | 349.56 | 11-13 | 1 | 261.30 |
| 4-8 | 2 | 416.14 | 12-14 | 2 | 348.37 |
| 5-6 | 1 | 415.00 | 12-16 | 2 | 406.46 |
| 5-7 | 1 | 435.50 | 13-16 | 2 | 417.27 |
| 5-10 | 1 | 376.35 | 14-15 | 1 | 458.18 |
| 6-9 | 1 | 316.35 | 14-17 | 1 | 403.64 |
| 7-11 | 1 | 387.10 | 15-17 | 2 | 402.16 |

The system incorporates all long transmission lines connecting the buses, facilitating efficient power transmission over long distances for such a voltage level. Notably, the length of the transmission lines varies, ranging from 261.30 km for line 11–13 to 458.18 km for line 14–17. This synthetic power network is specifically designed for transmission expansion planning purposes.

### *B. Generation and Load Information*

Table II presents comprehensive details on generation, load, and the required shunt capacitors during peak loading conditions. Bus 1 serves as the slack bus, with a voltage magnitude of $|V_1|$=1.05 p.u. and a voltage angle of $\delta_1$=0. Table II also outlines the voltage setting of other generators at peak load and their respective active power generation capacity. The reactive power generation limit in the synchronous generator is mainly defined by three conditions: thermal limit, voltage stability, and system stability. Generating excessive reactive power can lead to stator winding overheating due to increased currents, while over-excitation results in additional rotor heating. The second issue is that generating too much reactive power can lead to high system voltages, posing over-voltage risks, while absorbing too much reactive power can cause under-voltage, risking system destabilization. The other important issue is operating too far into the leading power factor region can weaken the magnetic coupling between the rotor and stator which increases the risk of instability Therefore, the reactive power generation limit is considered as $0.6P_{gi}$ when lagging power factor operating and $-0.3P_{gi}$ when operating at leading power factor. The system encompasses a total of 16 loads connected to each generation and load bus, excluding the slack bus. Each load operates with a lagging power factor of 0.9. During peak loading conditions, a total of 1200 Mvar fixed shunt capacitors are required to be connected to various buses within the system.

### *C. Transmission line configuration and line parameters*

The transmission line setup for the proposed 500 kV test system, depicted in Fig. 2, adopts horizontal arrangements with a phase spacing of 12.3 m, ensuring that the sub-conductors are positioned at 24 m above the ground level. Each phase comprises four sub-conductors bundle arranged in a circular formation, with a line spacing of 0.45 m. This tower configuration for a 500 kV transmission line is sourced from reference [32]. The designated sub-conductors, Macaw conductors, possess an outer diameter of 1.045 inches and a current carrying capacity of 870 Amperes. In a double-circuit line, each circuit maintains the same line configuration. For example, the two-line connecting buses 1 and 2 are positioned on two distinct towers, each with an identical configuration, as depicted in Fig. 2.

TABLE II
GENERATION, LOAD, AND SHUNT COMPENSATION AT PEAK LOAD

| Bus | \|V\| (p.u) | Bus Type | $P_g$ (MW) | $Q_{gmin}$ (Mvar) | $Q_{gmax}$ (Mvar) | $P_L$ (MW) | $Q_L$ (MW) | Shunt Capacitor |
|---|---|---|---|---|---|---|---|---|
| 1 | 1.05 | Slack | | | | | | |
| 2 | - | PQ | - | - | - | 1900.00 | 920.21 | 100 Mvar |
| 3 | 1.03 | PV | 3600 | -1080 | 2160 | 1750.00 | 847.56 | - |
| 4 | - | PQ | - | - | - | 1850.00 | 896.00 | 100 Mvar |
| 5 | - | PQ | - | - | - | 1600.00 | 774.92 | 150 Mvar |
| 6 | 1.03 | PV | 3600 | -1080 | 2160 | 1700.00 | 823.34 | - |
| 7 | - | PQ | - | - | - | 1900.00 | 920.21 | - |
| 8 | 1.04 | PV | 3600 | -1080 | 2160 | 1600.00 | 774.92 | - |
| 9 | - | PQ | - | - | - | 2000.00 | 968.64 | 450 Mvar |
| 10 | 1.03 | PV | 3600 | -1080 | 2160 | 1700.00 | 823.34 | - |
| 11 | - | PQ | - | - | - | 1800.00 | 871.77 | 200 Mvar |
| 12 | 1.05 | PV | 3600 | -1080 | 2160 | 1600.00 | 774.92 | - |
| 13 | 1.05 | PV | 3600 | -1080 | 2160 | 1800.00 | 871.77 | - |
| 14 | - | PQ | - | - | - | 2300.00 | 1113.94 | 50 Mvar |
| 15 | 1.00 | PV | 3500 | -1050 | 2100 | 1700.00 | 823.35 | - |
| 16 | - | PQ | - | - | - | 1750.00 | 847.56 | - |
| 17 | - | PQ | - | - | - | 1150.00 | 556.97 | 150 Mvar |

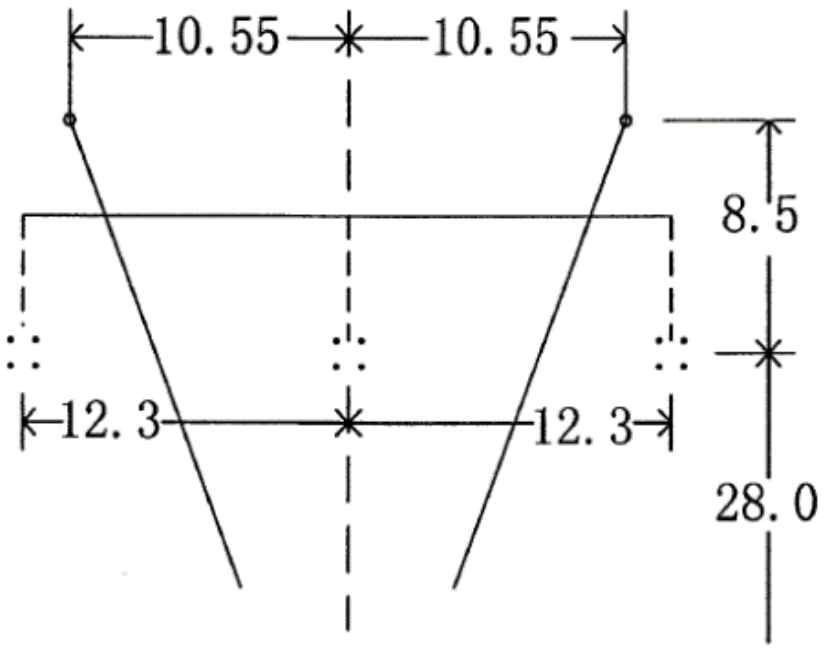

Fig. 2. 500 kV transmission line configuration used in the test system.

The line parameters of transmission lines encompass a range of electrical attributes delineating the line's performance in transmitting electrical energy. For a given number of bundle conductors per phase denoted as $b$, these parameters can be computed as

$$R_{eq} = \frac{R}{b} \quad \Omega/\text{km} \tag{1}$$

$$x = 2\pi f \times 2 \times 10^{-7} \ln\left(\frac{GMD}{r'}\right) \quad \Omega/\text{km} \tag{2}$$

$$y = 2\pi f \frac{2\pi\varepsilon_0}{\ln\left(\frac{GMD}{r_0}\right)} \quad S/km \tag{3}$$

where $f$ represents the system frequency, $\varepsilon_0$ denotes the permittivity of free space, and $GMD$ stands for the geometric mean distance. When dealing with bundled conductors, the parameters $r'$ and $r_0$ should be substituted with the equivalent bundle radii for inductance calculations and capacitance calculations.

*C. Modeling of transmission line*

When dealing with long transmission lines, relying exclusively on the multiplication of per-unit-length line parameters by the transmission line length for calculating total line impedance and shunt admittance can result in inaccuracies. As the transmission line's length increases, the margin of error in estimating these parameters through simple distance scaling also increases. Table III and Fig. 3 illustrate the specific percentage disparities in resistance, reactance, and susceptance resulting from this direct scaling method compared to the actual parameters. These findings highlight a significant increase in the disparity of resistance and inductive reactance, particularly notable for transmission lines exceeding 150 miles in length.

Thus, achieving precise modeling of long transmission lines requires ensuring consistent distribution of line parameters across their lengths. Directly scrutinizing these lines with distributed parameter models entails significant computational complexity. The $\pi$ model offers a streamlined approach by consolidating distributed parameters into lumped elements, thereby presenting a simplified yet precise portrayal of long transmission lines. This equivalent $\pi$ model of long transmission lines is depicted in Fig. 4. The equivalent series impedance is denoted as $Z'$ and the equivalent shunt admittance is denoted as $Y'$.

TABLE III

DIFFERENCES IN LINE PARAMETERS WITH DISTRIBUTED AND LUMPED MODELING

| Length (miles) | Difference, % | | |
|---|---|---|---|
| | ΔR | ΔX | ΔB |
| 10 | 0 | 0 | 0 |
| 50 | 0.35 | 0.18 | 0 |
| 100 | 1.43 | 0.71 | 0.35 |
| 200 | 5.90 | 2.87 | 1.42 |
| 300 | 14.09 | 6.62 | 3.20 |
| 400 | 27.33 | 12.21 | 5.71 |
| 500 | 48.34 | 20.03 | 8.98 |

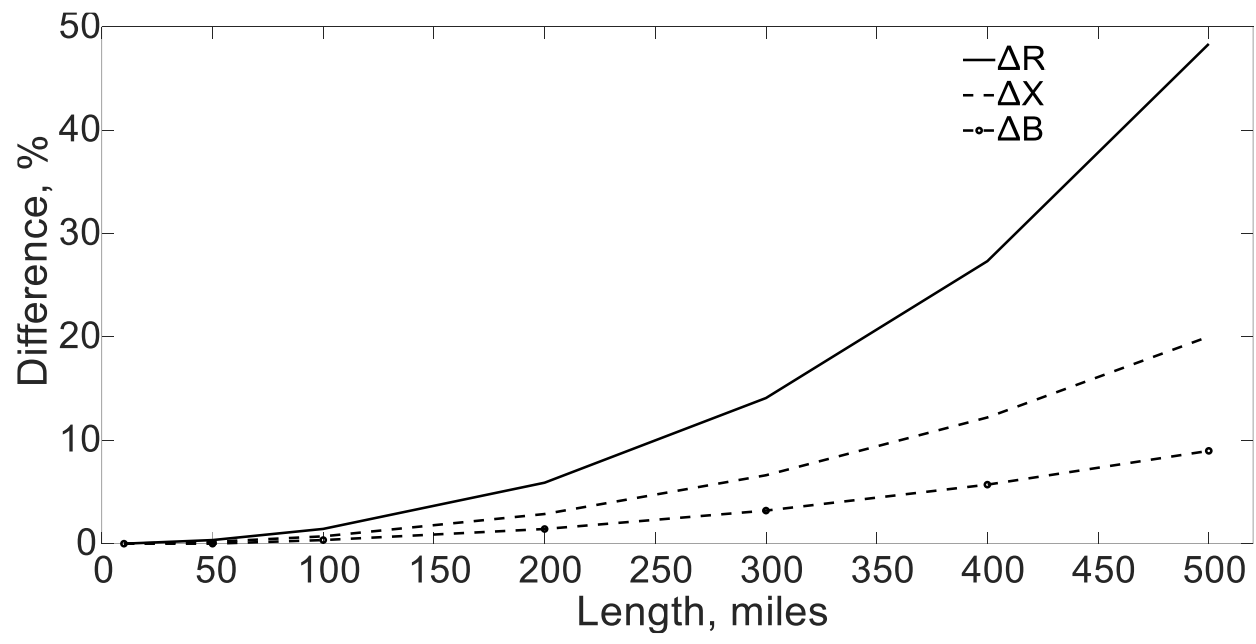

Fig. 3. Difference in line parameter with distributed and lumped modeling.

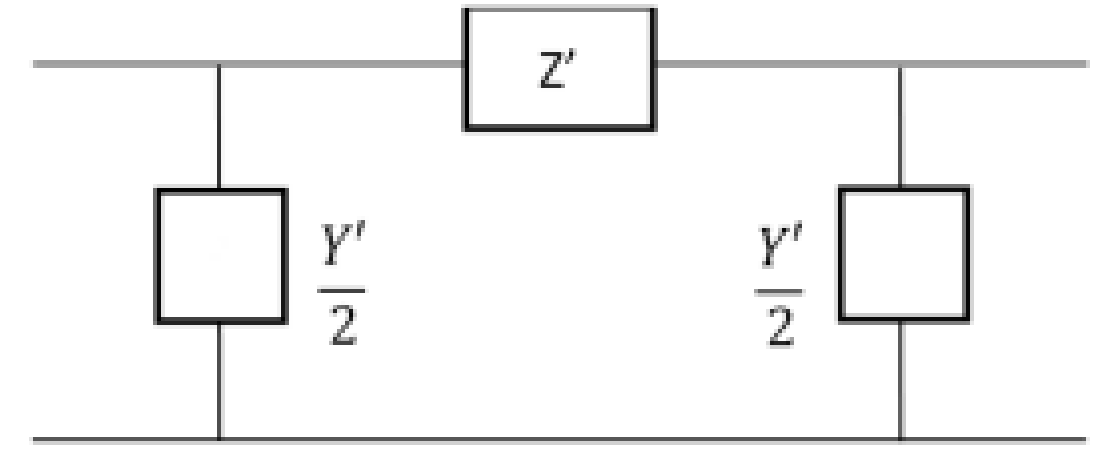

Fig. 4. Equivalent $\pi$ model of a long transmission line

The equivalent series impedance and shunt charging admittance for this modeling can be computed as:

$$Z' = zl\frac{sinh(\gamma l)}{\gamma l} \tag{4}$$

$$Y' = yl\frac{tanh(\gamma l/2)}{(\gamma l/2)} \tag{5}$$

where $\gamma$ represents the propagation constant defined as $\gamma = \sqrt{zy}$. The $l$ denotes the length of the transmission line, while $z$ and $y$ are series impedance and shunt admittance per unit length of the transmission line.

The thermal limit of a transmission line, denoted as $MVA_{max}$, can be determined through a calculation involving the line voltage, $V_{line}$, the maximum permissible current within the conductor, $I_{max}$, and the number of sub-conductors per phase, as depicted in Eq. (6).

$$MVA_{max} = \sqrt{3} \times V_{\text{line}} \times I_{max} \times b \tag{6}$$

The line parameters and the maximum line loading limit, $S_{in}^{max}$, considered as 80% of the thermal limit, for the considered transmission line structure is given in Table IV. Note that thermal limit is classified into static line rating (SLR) and dynamic thermal rating (DTR) [33, 34]. In the proposed test system, all transmission lines are long, with lengths ranging from 262 km to 458 km. Under normal and single contingency conditions, line loading remains well below the static thermal limit rating (SLR) of the transmission lines.

TABLE IV
LINE PARAMETERS

| kV | ACSR Conductor | Line Constraints | | | $S_{in}^{max}$ |
|---|---|---|---|---|---|
| | | R (Ω/km) | L (mH/km) | C (nF/km) | (MVA limit) |
| 500 | Macaw, 4/bundle | 0.0228 | 0.878 | 12.975 | 2411 |

## III. FORMULATION OF LOAD FLOW ANALYSIS

Power flow analysis plays a pivotal role in ensuring the efficient operation of power systems. Through precise modeling and simulation of power flow within a network, potential issues such as overloading, voltage violations, and the area of congestion can be anticipated. This predictive capability enables proactive measures to be implemented, such as adjusting generation levels, redistributing loads, or enhancing infrastructure, thereby maintaining system stability and reliability. Furthermore, power flow analysis informs strategic planning for system expansion and upgrades by offering valuable insights into resource allocation for maximum effectiveness. To ascertain the technical viability of the proposed test system, a detailed load flow analysis needs to be carried out under normal and all single contingency conditions at multiple loading scenarios.

The generalized equations for the load flow analysis are illustrated in Eqs (7)-(10).

$$I = Y_{bus}V \tag{7}$$

$$P_i + jQ_i = V_i I_i^* \tag{8}$$

$$P_i = |V_i| \sum_{n=1}^{N} |V_n||Y_{in}|\cos(\theta_{in} - \delta_i + \delta_n) \tag{9}$$

$$Q_i = -|V_i| \sum_{n=1}^{n} |V_n||Y_{in}|\sin(\theta_{in} - \delta_i + \delta_n) \tag{10}$$

where, $|Y_{in}|$ and $\theta_{in}$ are the magnitude and angle, respectively, of an element in the bus admittance matrix, $|V|$ and $\delta$ denotes the magnitude and angle of the bus voltage, $I$ represents the bus-injected current, and $P$ and $Q$ are the injected active and reactive power into the buses, respectively.

The power flow problem is subject to the following constraints.

$$\text{Normal condition: } 0.95 \leq |V_i| \leq 1.05\ p.u. \tag{11}$$

$$\text{Contingency condition: } 0.90 \leq |V_i| \leq 1.05\ p.u. \tag{12}$$

$$-0.3P_{gi} \leq Q_{gi} \leq 0.6P_{gi} \tag{13}$$

$$S_{in} \leq S_{in}^{max} \tag{14}$$

Eqs. (11) and (12) are the constraints aimed at maintaining the voltage magnitude of all buses within acceptable limits during normal operating conditions and all single contingency scenarios, respectively. Eq. (13) ensures the reactive power generation by all generator units connected to the PV bus remains within the specified limit, and this is the practical consideration for the operation of synchronous generators. The constraint outlined in Eq. (14) ensures that power flowing through the transmission line connecting two buses does not surpass their maximum capacity, determined by the thermal limit.

Contingency scenarios are common in power systems, emphasizing the need for the system to remain operational under all single contingency conditions to ensure grid resilience, reliability, and robustness. In transmission expansion planning (TEP) analysis, typical constraints include the outage of transmission lines and transformers. This test system is designed to remain functional under such conditions, specifically considering line outages as contingency components. For this analysis, load flow equations (7)-(10) and constraints (12)-(14) are applicable. The voltage magnitude requirement at each bus is adjusted to be above 0.90 p.u. rather than the usual 0.95 p.u. as specified in Eq. (12). This 0.90 p.u. is the minimum threshold during contingencies is commonly accepted by utilities in practical TEP studies.

## IV. POWER FLOW RESULT AND ANALYSIS

The power flow analysis was conducted using the Newton-Raphson method in the PSS/E 35.4 software. Generation and load data presented in Table II correspond to peak loading conditions. Since the power system does not operate continuously at peak load, it is essential to analyze various loading conditions to ensure reliability and stability under different scenarios. For TEP studies, loadings in different seasons during a year are often considered where peak load is summer, dominant loading is for spring and fall, and light load is for winter; this is often for hot countries where electricity is used for cooling during summer while having a moderate winter. Analysis was performed for normal and all single contingency conditions for three different loading conditions, and the results are detailed in this section.

### *A. Peak Load*

#### *A.1. Normal Operating Conditions*

Fig. 5 illustrates the detailed power flow outcome for the 17-bus test system during normal operating condition. In this graphical depiction of power flow, numerical values on either side of each transmission line, generating unit, load, and shunt reactor indicate the values of active and reactive powers. Positive and negative numerical values denote the power output and input to the respective buses where transmission lines are connected, and opposite polarity applies to generating units and fixed shunt compensator devices. The numerical values associated with each bus signify the bus identification number, voltage magnitude in per unit, and its angle. The summarized result of this power flow analysis is presented in Table V. The result indicates that per unit voltage at all buses and reactive power generations at the generating unit connected to PV buses are within the defined range. Additionally, three lines – line 5 to 6, line 2 to 3, and line 6 to 9 – are identified as lines carrying maximum power, with loading percentages of 32.50%, 32.50%, and 31.74%, respectively, all well below their maximum capacity. The total power loss in the power system network at this condition was 321.2 MW.

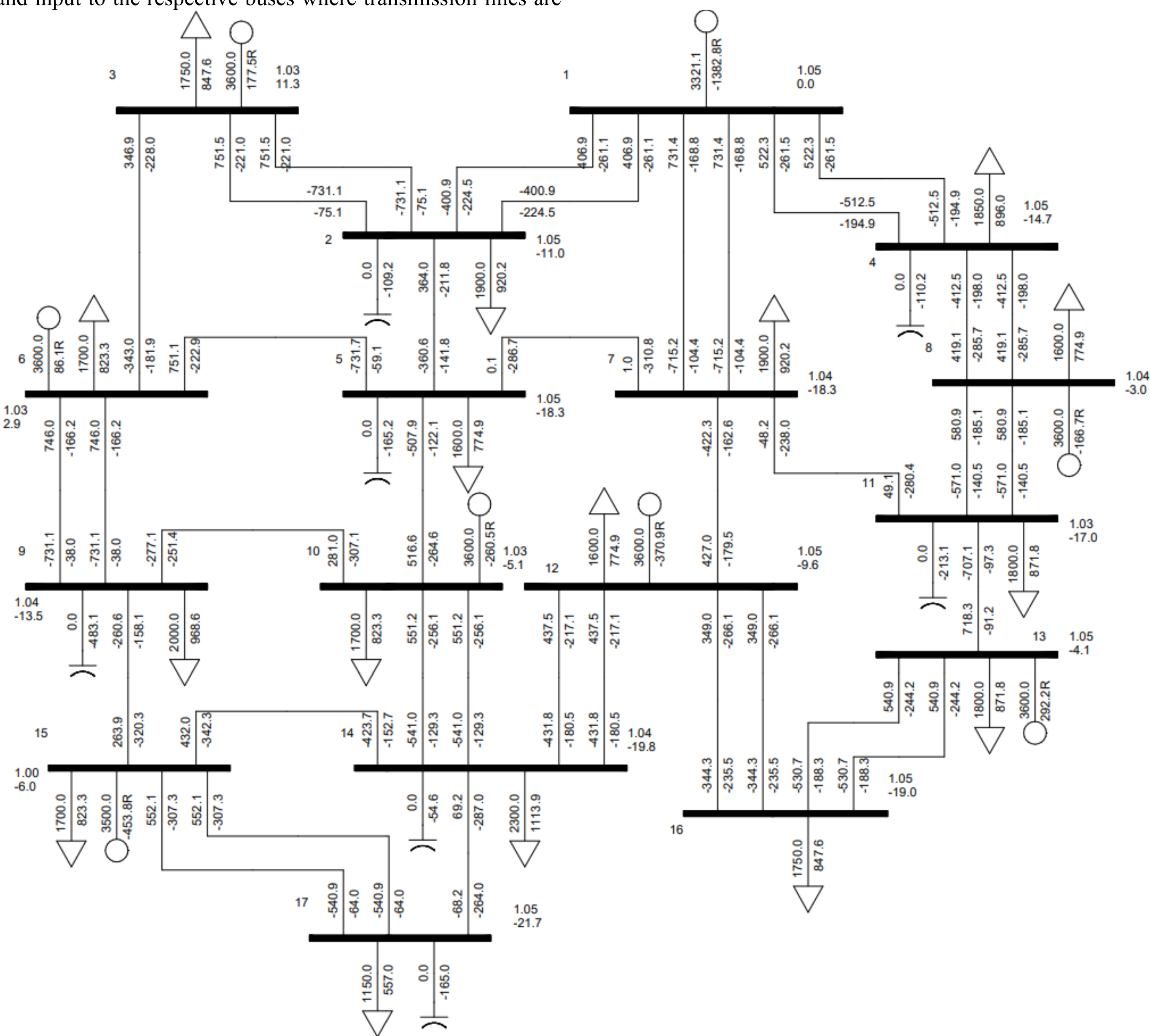

Fig. 5. Power flow for the test system at peak load under normal operating condition.

*A.2.* Contingency conditions

The test system's performance under peak load during contingency was evaluated by deactivating individual transmission lines and conducting load flow analyses for each outage. The summarized load flow result for each specific contingency scenario is compiled and presented in Table VI. Each row of the table corresponds to a different contingency situation, displaying the lowest voltage occurring at a bus and the highest loading line, along with their respective values. A notable severe contingency occurred when one of the lines connecting bus 15 to bus 17 was not operational. Under this contingency, the lowest voltage was recorded at bus 17 having a voltage magnitude of 0.907 p.u. Additionally, a maximum loading of 53.40% was observed in line 6- 9 when another line 6-9 was not in service. It is noteworthy that the outcomes for every single contingency condition adhered to the voltage limit, line loading limit, and reactive power generation limits as outlined in Eqs. (12) to (14). This demonstrates the test system's capability to effectively operate under all single contingency scenarios at peak loading condition.

*B. Dominant Load*

The dominant loading condition was considered as 60% of the peak loading conditions. When examining the system under this condition, the voltage across all generating units was set to 1.0 p.u. Moreover, a total cumulative shunt reactor capacity of 2350 Mvar was required to be connected across different buses. Table VII provides details on the allocated buses and their respective shunt reactor capacities.

TABLE V
SUMMARIZED LOAD FLOW ANALYSIS RESULT AT PEAK LOAD UNDER NORMAL CONDITION

| Bus # | Voltage | | Generation | |
|---|---|---|---|---|
| | $\|V\|$ $p.u.$ | $\delta$ (deg.) | $P_g$ (MW) | $Q_g$ (Mvar) |
| 1 | 1.050 | 0.00 | 3321.2 | -1382.6 |
| 2 | 1.045 | -11.04 | 0.0 | 0.0 |
| 3 | 1.030 | 11.35 | 3600.0 | 177.8 |
| 4 | 1.050 | -14.69 | 0.0 | 0.0 |
| 5 | 1.049 | -18.32 | 0.0 | 0.0 |
| 6 | 1.030 | 3.52 | 3400.0 | 130.8 |
| 7 | 1.042 | -18.29 | 0.0 | 0.0 |
| 8 | 1.040 | -3.05 | 3600.0 | -166.7 |
| 9 | 1.028 | -13.62 | 0.0 | 0.0 |
| 10 | 1.030 | -5.07 | 3600.0 | -244.6 |
| 11 | 1.032 | -16.97 | 0.0 | 0.0 |
| 12 | 1.050 | -9.63 | 3600.0 | -370.9 |
| 13 | 1.050 | -4.08 | 3600.0 | 292.2 |
| 14 | 1.045 | -1985 | 0.0 | 0.0 |
| 15 | 1.000 | -6.06 | 3500.0 | -436.8 |
| 16 | 1.046 | -19.00 | 0.0 | 0.0 |
| 17 | 1.049 | -21.71 | 0.0 | 0.0 |

TABLE VI
SUMMARIZED LOAD FLOW ANALYSIS RESULT AT PEAK LOAD UNDER ALL SINGLE CONTINGENCIES

| Line outage | Lowest Voltage | | The Highest Line Loading | |
|---|---|---|---|---|
| | $\|V\|$ $p.u.$ | Bus# | %loading | Line |
| 1–2 (1 line) | 0.972 | 2 | 33.63% | 1–7 |
| 1–4 (1 line) | 0.932 | 4 | 36.33% | 1–4 |
| 1–7 (1 line) | 0.969 | 7 | 45.56% | 1–7 |
| 2–3 (1 line) | 0.949 | 2 | 49.76% | 2–3 |
| 2–5 | 0.969 | 5 | 36.00% | 1–7 |
| 3–6 | 1.000 | 15 | 38.59% | 2–3 |
| 4–8 (1 line) | 0.947 | 4 | 32.56% | 6–9 |
| 5–6 | 0.935 | 5 | 39.85% | 6–9 |
| 5–7 | 0.972 | 5 | 31.64% | 6–9 |
| 5–10 | 0.938 | 5 | 39.05% | 5–6 |
| 6–9 (1 line) | 0.918 | 9 | 53.40% | 6–9 |
| 7–11 | 0.984 | 11 | 32.50% | 2–3 |
| 7–12 | 0.978 | 7 | 38.05% | 11–13 |
| 8–11 (1 line) | 0.956 | 11 | 39.50% | 8–11 |
| 9–10 | 0.946 | 9 | 34.16% | 6–9 |
| 9–15 | 0.978 | 9 | 34.04% | 6–9 |
| 10–14 (1 line) | 0.996 | 14 | 35.75% | 10–14 |
| 11–13 | 0.953 | 11 | 37.62% | 7–12 |
| 12–14 (1 line) | 0.994 | 14 | 32.68% | 6–9 |
| 12–16 (1 line) | 0.954 | 16 | 32.67% | 5–6 |
| 13–16 (1 line) | 0.940 | 16 | 37.89% | 11–13 |
| 14–15 | 1.000 | 15 | 34.27% | 5–6 |
| 14–17 | 0.930 | 17 | 32.65% | 2–3 |
| 15–17 (1 line) | 0.907 | 17 | 36.54% | 15–17 |

*B.1. Normal Operating Condition*

Fig. 6 shows a comprehensive load flow analysis conducted on the test system under the dominant load during normal operating conditions. The summarized result of this analysis is presented in Table VII. The findings reveal the adherence of both the voltage levels at all buses and the reactive power generation from all generating units connected to PV buses to their defined thresholds, as expressed in Eqs. (11) and (13). The three highest loading lines in this condition are line 2 – 3 with a loading percentage of 23.45%, line 5 – 6 with 23.21%, and line 6 – 9 with 21.00 %, all well below their limits, as defined in Eq. (14).

TABLE VII
SUMMARIZED POWER FLOW RESULT OF THE TEST SYSTEM AT DOMINANT LOAD UNDER NORMAL OPERATING CONDITION

| Bus # | Voltage | | Generation | |
|---|---|---|---|---|
| | $\|V\|$ $p.u.$ | $\delta$ (deg.) | $P_g$ (MW) | $Q_g$ (Mvar) |
| 1 | 1.000 | 0.00 | 1945.3 | -2030.25 |
| 2 | 1.048 | -6.85 | 0.0 | 0.0 |
| 3 | 1.000 | 7.07 | 2160.0 | -383.29 |
| 4 | 1.044 | -9.27 | 0.0 | 0.0 |
| 5 | 1.050 | -11.19 | 0.0 | 0.0 |
| 6 | 1.000 | 1.95 | 2160.0 | -520.87 |
| 7 | 1.049 | -11.32 | 0.0 | 0.0 |
| 8 | 1.000 | -1.75 | 2160.0 | -594.42 |
| 9 | 1.024 | -8.29 | 0.0 | 0.0 |
| 10 | 1.000 | -2.87 | 2160.0 | -641.17 |
| 11 | 1.032 | -10.45 | 0.0 | 0.0 |
| 12 | 1.000 | -5.55 | 2160.0 | -642.18 |
| 13 | 1.000 | -2.13 | 2160.0 | -410.40 |
| 14 | 1.048 | -12.06 | 0.0 | 0.0 |
| 15 | 1.000 | -3.71 | 2100.0 | -587.97 |
| 16 | 1.045 | -11.63 | 0.0 | 0.0 |
| 17 | 1.044 | -13.12 | 0.0 | 0.0 |

*B.2. Contingency Condition*

In the dominant load, load flow analysis was carried out for each single line outage, akin to the contingency analysis conducted under peak loading conditions. The specific outcomes of each contingency condition are summarized in Table VIII. It was determined that the outage of line 5–6 constituted the most critical contingency for the dominant loading condition. Under this scenario, bus 5 recorded the minimum voltage, measuring 0.938 p.u. Despite the severity of this contingency, all buses maintained voltages well above their lower limit and within their higher limit. Furthermore, the other constraints, such as reactive power generation limits and line loading under contingencies, were found to meet specified thresholds.

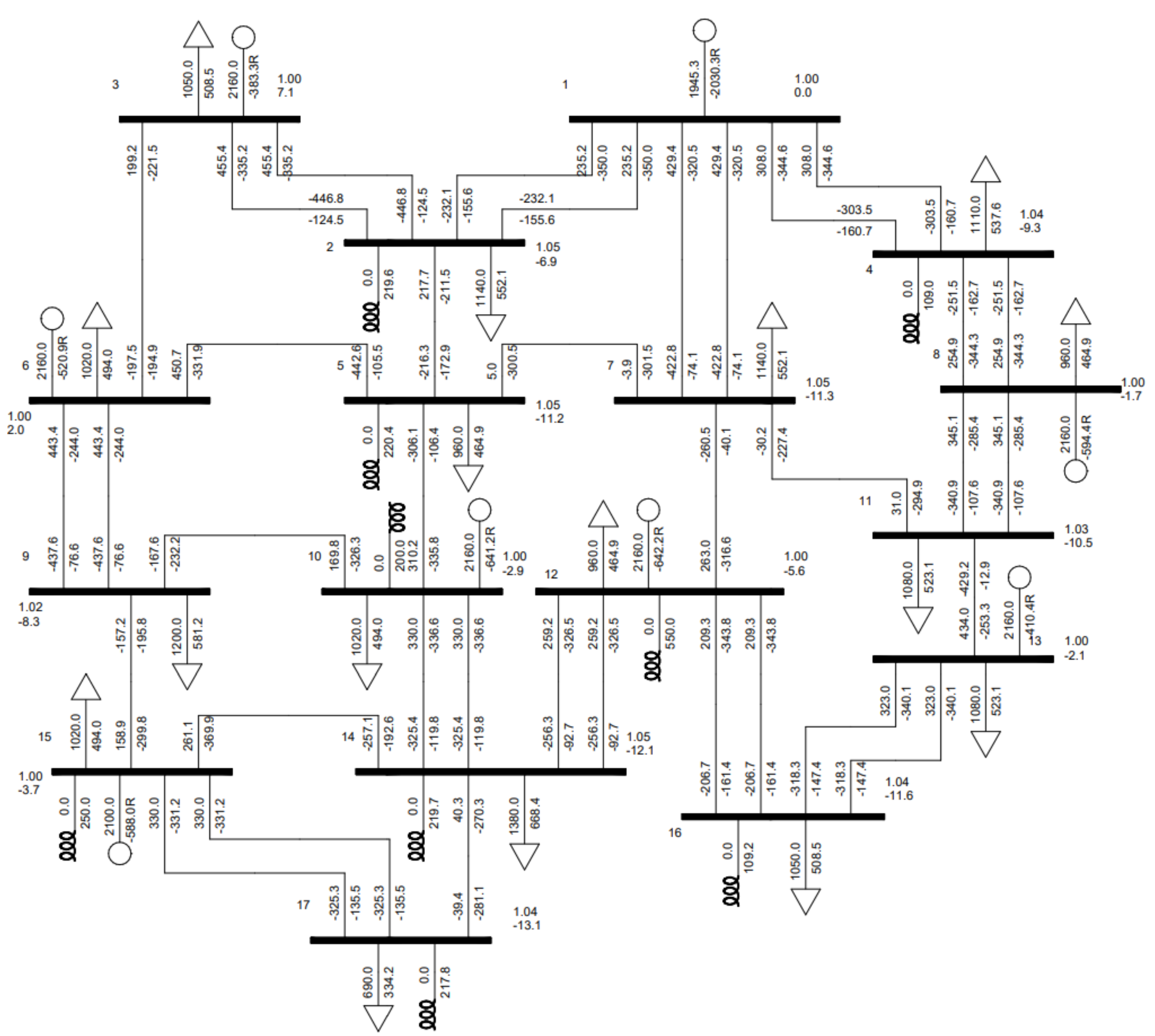

Fig. 6. Power flow result of the test system at dominant load under normal operating condition.

*C. Light Load: Normal and contingency operating conditions*

The condition of light loading was described as operating at 40% of the peak load condition. To ensure operational adequacy under this loading while adhering to technical specifications such as bus voltage and reactive power generation, a collective capacity of 6800 Mvar shunt reactors had to be incorporated across various buses. Information regarding the buses to which the shunt reactors were connected and their respective capacities can be found in Table IX. All generating units were configured to operate at a voltage of 1.0 p.u. A load flow analysis was carried out under normal operating conditions, and the comprehensive result is presented in Fig. 7. This result demonstrates that the per-unit voltage levels at each bus and the reactive power generation of generating units connected to PV buses are within their specified thresholds. Even under single contingency conditions, the minimum observed voltage remained comfortably above the threshold, noted at a value of 0.963 p.u. when line 14–17 was out of operation.

The results of the comprehensive load flow analysis conducted on the proposed test system under normal operating condition and considering all single contingencies across three different loading scenarios affirm adherence to the technical requirements outlined in Eqs. (11)–(14). This validation underscores the system's resilience in effectively handling a wide range of load fluctuations, spanning from the light load to peak load, across normal operations and all single contingency scenarios.

The system requires maximum shunt reactors under minimum loading conditions and maximum shunt capacitive compensation during peak loading. Therefore, the locations and capacities of shunt reactors at light loading condition and shunt capacitors at peak loading condition are determined and fixed as the base test system for further studies.

TABLE VIII
SUMMARIZED LOAD FLOW ANALYSIS RESULT AT DOMINANT LOAD UNDER ALL SINGLE CONTINGENCIES

| Line outage | Lowest Voltage | | The Highest Line Loading | |
|---|---|---|---|---|
| | $\|V\|$ $p.u.$ | Bus # | % loading | Line |
| 1–2 (1 line) | 1.000 | PV Bus | 23.20% | 1–7 |
| 1–4 (1 line) | 0.989 | 4 | 22.72% | 1–4 |
| 1–7 (1 line) | 1.000 | PV Bus | 28.26% | 1–7 |
| 2–3 (1 line) | 0.949 | PV Bus | 31.42% | 2–3 |
| 2–5 | 0.999 | 5 | 23.42% | 5–6 |
| 3–6 | 1.000 | PV Bus | 26.20% | 2–3 |
| 4–8 (1 line) | 0.989 | 4 | 23.44% | 1–7 |
| 5–6 | 0.935 | 5 | 25.66% | 2–3 |
| 5–7 | 0.989 | 5 | 22.62% | 2–3 |
| 5–10 | 0.938 | PV Bus | 25.33% | 5–6 |
| 6–9 (1 line) | 0.982 | 9 | 31.22% | 6–9 |
| 7–11 | 0.986 | 11 | 23.30% | 2–3 |
| 7–12 | 1.000 | PV Bus | 24.70% | 11–13 |
| 8–11 (1 line) | 0.998 | 11 | 25.22% | 8–11 |
| 9–10 | 0.975 | 9 | 21.42% | 6–9 |
| 9–15 | 0.981 | 9 | 23.06% | 2–3 |
| 10–14 (1 line) | 1.000 | PV Bus | 23.98% | 10–14 |
| 11–13 | 1.000 | PV Bus | 25.13% | 7–12 |
| 12–14 (1 line) | 0.994 | PV Bus | 23.13% | 2–3 |
| 12–16 (1 line) | 0.999 | 16 | 23.53% | 2–3 |
| 13–16 (1 line) | 0.995 | 16 | 24.42% | 11–13 |
| 14–15 | 1.000 | PV Bus | 24.08% | 5–6 |
| 14–17 | 0.955 | 17 | 23.42% | 2–3 |
| 15–17 (1 line) | 0.969 | 17 | 22.86% | 15–17 |

## V. CASE STUDIES FOR TRANSMISSION EXPANSION PLANNING

Based on the proposed test system presented in this paper, six TEP case studies were conducted, aiming to transmit power to a new location identified as bus 18. Bus 18 was positioned at a distance of 324.54 km from bus 16 and 341.84 km from bus 17. These TEP analyses were performed with varying numbers of line connections between bus 16 and bus 17 to bus 18. The objective was to assess the maximum power deliverable to the load at bus 18 in each TEP case, ensuring compliance with all technical requirements under normal operating conditions and all single contingency scenarios across peak load, dominant load, and light loading conditions. The same transmission line configuration as presented in Section II was used for the TEP studies. When the system configuration changed after TEP, the reactive power compensation requirements at different loading conditions at new line-connected buses also changed. Therefore, the locations of shunt compensation remain unchanged as those of the base system and the system after each TEP case study. However, the capacities of new line-connected buses needed adjustment.

### *A. Case I: Two/two-line connections from buses 16 and 17 to bus 18*

In the TEP case with two/two-line connections from buses 16 and 17 to bus 18, it was able to deliver a maximum load of 1115 MW at bus 18 with a 0.9 lagging power factor. This setup effectively met all technical prerequisites for normal and all single contingency conditions, across three different loading scenarios. The total capacity of shunt compensation at peak load is 1100 Mvar shunt capacitors and 550 Mvar shunt reactors. Values with the notation '(Cap.)' in the Mvar row signify the shunt capacitor Mvar capacity, while the values without '(Cap.)' notation indicate the shunt inductor Mvar capacity. The detailed power flow result for this TEP case at peak load under normal operating conditions is depicted in Fig. 8. The results indicate that the operational requirements for normal operating condition, as specified in Eqs. (11), (13), and (14), are satisfactorily met. Moreover, even under contingency scenarios, the minimum voltage, observed at bus 18 with a magnitude of 0.900 p.u., remained in accordance with the prescribed threshold outlined in Eq. (12). Furthermore, the highest line loading during a contingency occurred in line 1–7 when another line 1 – 7 was inactive.

In order to uphold the technical viability of the test system for dominant and light loading conditions under normal and all single contingency scenarios, adjustments to shunt reactors were deemed necessary. A cumulative value of 2700 Mvar shunt reactor was required for the dominant loading condition. During contingency situations under dominant loading, the minimum voltage was observed at bus 18 (0.921 p.u.) when a line connecting buses 16 and 18 went out of service, and the maximum loaded line occurred in line 1 – 7 (45.12%) when the other line 1 – 7 was out of operation. Similarly, for light loading conditions, a total of 7200 Mvar shunt reactors needed to be incorporated into the system to ensure satisfactory operation under normal and all single contingency circumstances. During contingencies in light loading, the lowest voltage was observed at bus 18 (0.930 p.u.) when a line connected to buses 16 and 18 was not operational. Details regarding the shunt reactors connected to buses and their respective capacities for different loading conditions are provided in Table X.

TABLE IX
INFORMATION ON THE REQUIRED SHUNT REACTOR AT LIGHT LOAD

| Dominant Loading Condition | | | | | | | | | | | | | | | | |
|---|---|---|---|---|---|---|---|---|---|---|---|---|---|---|---|---|
| Bus | 2 | 3 | 4 | 5 | 6 | 7 | 8 | 9 | 10 | 11 | 12 | 13 | 14 | 15 | 16 | 17 |
| Mvar | 200 | - | 100 | 200 | - | 150 | 200 | - | 200 | - | 550 | - | 200 | 250 | 100 | 200 |
| Light Loading Condition | | | | | | | | | | | | | | | | |
| Bus | 2 | 3 | 4 | 5 | 6 | 7 | 8 | 9 | 10 | 11 | 12 | 13 | 14 | 15 | 16 | 17 |
| Mvar | 450 | 200 | 300 | 400 | 450 | 400 | 650 | 100 | 700 | 100 | 1000 | 250 | 450 | 700 | 300 | 350 |

Fig. 7. Power flow result of the test system at light load under normal operating condition.

TABLE X
INFORMATION OF REQUIRED SHUNT COMPENSATION AT PLEAK LOAD, DOMINANT LOAD, AND LIGHT LOAD FOR THE TEP CASE I

| Peak Loading Condition | | | | | | | | | | | | | | | | | |
|---|---|---|---|---|---|---|---|---|---|---|---|---|---|---|---|---|---|
| Bus | 2 | 3 | 4 | 5 | 6 | 7 | 8 | 9 | 10 | 11 | 12 | 13 | 14 | 15 | 16 | 17 | 18 |
| Mvar | 100 (Cap.) | - | 100 (Cap.) | 150 (Cap.) | - | - | - | 450 (cap.) | - | 200 (Cap.) | - | - | 50 (Cap.) | - | 200 | 150 | 200 |
| Dominant Loading Condition | | | | | | | | | | | | | | | | | |

| Bus | 2 | 3 | 4 | 5 | 6 | 7 | 8 | 9 | 10 | 11 | 12 | 13 | 14 | 15 | 16 | 17 | 18 |
|---|---|---|---|---|---|---|---|---|---|---|---|---|---|---|---|---|---|
| Mvar | 150 | - | 50 | 200 | - | - | 150 | - | 200 | - | 550 | - | 150 | 200 | 300 | 500 | 250 |
| Light Loading Condition | | | | | | | | | | | | | | | | | |
| Bus | 2 | 3 | 4 | 5 | 6 | 7 | 8 | 9 | 10 | 11 | 12 | 13 | 14 | 15 | 16 | 17 | 18 |
| Mvar | 400 | 200 | 300 | 400 | 450 | 300 | 600 | 50 | 650 | 100 | 950 | 200 | 450 | 700 | 550 | 650 | 250 |

*B. Summarized result of six TEP cases*

The detailed analysis of TEP with different numbers and combinations of line connections from buses 16 and 17 to bus 18, along with the corresponding maximum power supply to bus 18, system power loss at peak loading conditions, and the total capacity of required shunt compensation for each loading condition (peak load, dominant load, and light loading conditions) for each TEP case was meticulously evaluated, as illustrated in Case I presented in Section V(A), and summarized in Table XI. The results revealed that the maximum power delivered through a total of two-line combinations from either bus was 130 MW, while the maximum power delivered from three-line combinations from either bus was 620 MW, and from four-line combinations from either bus was 1115 MW. In all TEP cases, technically viable system operation under normal and all single contingencies for all three loading scenarios was ensured. The maximum capacity of shunt reactors needed to be added to the system ranged from 400 Mvar to 1250 Mvar more compared to the shunt reactors needed in the base test system, ensuring satisfactory operation and meeting all technical requirements outlined in Eqs. (11) – (14). For instance, Case I, with two-line connections from buses 16 and 17 to bus 18, required an additional shunt reactor with a capacity of 400 Mvar, while Case III, with one line connection from bus 16 to bus 18 and three line connections from bus 17 to 18 necessitated more additional shunt reactors with an extra capacity of 1250 Mvar compare to the base test system.

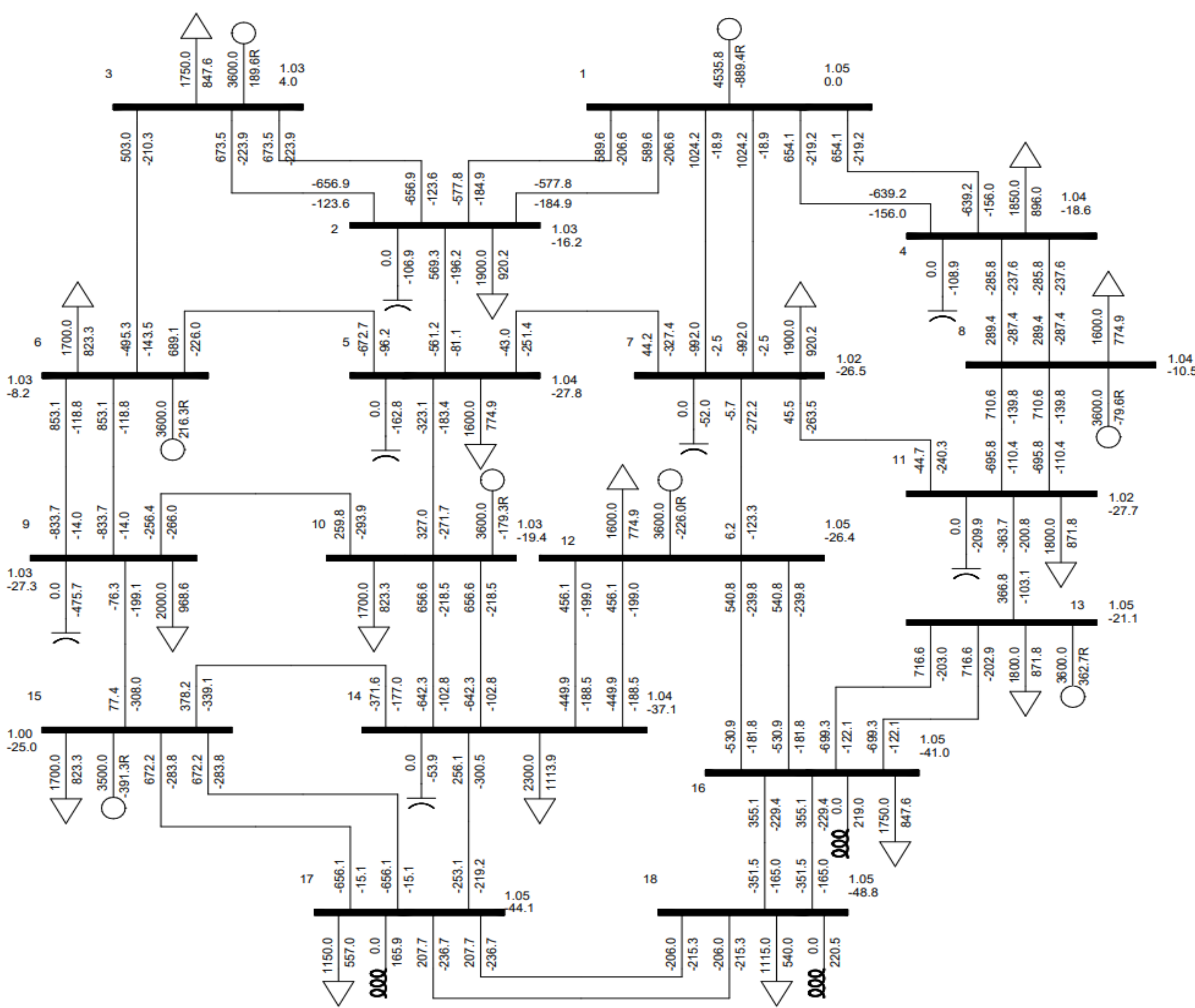

Fig. 8. Power flow result at peak load under normal operating conditions after TEP with Case I: Two/two-line connections from bus 16 and 17 to bus 18.

### C. Cost Analysis of TEP Cases

To calculate the cost of each TEP case, it is necessary to consider the cost of an additional 500 kV transmission line, the cost of the shunt reactor, the cost of the bay, and the cost of energy loss. Hence, the total cost of TEP can be calculated as:

$$C_{TEP} = l_{line}C_{line} + n_{reactor}C_{reactor} + n_{bays}C_{bays} + e_{loss}C_{loss} \quad (15)$$

where $l_{line}$ is the length of added line, $C_{line}$ is the cost per km of the 500 kV transmission line; $n_{reactor}$ is the additional capacity of shunt reactor in Mvar, and $C_{reactor}$ is the cost of shunt reactor per Mvar; $n_{bays}$ is the number of additional bays, and $C_{bays}$ is the cost per unit bay; $e_{loss}$ is the additional energy loss in MWh due to TEP.

Valuable insights into these cost components are outlined in the 2023 Midcontinent Independent System Operator (MISO) report on the Transmission Cost Estimation Guide [35], which offers detailed on the average transmission line costs, expenses associated with the addition of bay for establishing new line connections, and the unit costs associated to grid-supporting apparatus such as shunt reactors or shunt capacitors across diverse voltage levels, including the 500 kV level. According to the aforementioned report, the average cost of a single circuit 500 kV transmission line in various states across the United States stands at $2.672 million/km. Furthermore, the cost linked with integrating a bay into a 500 kV substation, equipped with a breaker and half-bus characteristics to facilitate a new line connection, amounts to $7.0 million. Additionally, the report stipulates that the cost per Mvar of the shunt reactor is estimated at $23,625. It is noteworthy that connecting the shunt compensating devices to bus 18, which serves as the new bus after TEP, necessitates an additional bay.

The cost of the power losses must also be accounted for in transmission expansion planning. These losses in the system will be supplied through the generation units, and the cost of losses can be calculated in terms of generation cost, equal to the amount of power loss. Reference [36] indicates that for natural gas-fired combined cycle power plants in the U.S., the average capital cost per MW of generation units is 1.04 million. Additionally, it has been stated in reference [37] that the fixed operational and maintenance (O&M) cost for a natural gas combined cycle power plant is $30,000 per MWh, and the variable O&M cost is $1.92 per MWh. Hence, the total operation and maintenance cost per year for 1MW power generation would amount to $0.046819 million. In the year 2023, the average cost of natural gas stands at $2.665/MMbtu [38], with a heat rate of 6.30MMbtu/MWh [37]. Therefore, the annual cost of fuel for a 1 MW power plant totals $0.147076 million. Considering the life span of the transmission line to 30 years, and assuming the constant O&M cost and fuel cost throughout the considered period, the cost of each TEP case was evaluated and is presented in Table XII. The cost analysis results reveal that the minimum average cost per MW power transfer to bus 18 is for Case I, with a two/two-line connection from buses 16 and 17 to bus 18, while the maximum average cost per MW power transfer to the bus is for Case VI, with a one/one line connection from buses 16 and 17 to bus 18.

TABLE XI

SUMMARIZED RESULT OF SIX TEP CASE STUDIES

| TEP Case | No. of line from 16 to 18 | No. of line from 17 to 18 | Max. power deliver to bus 18 | Power Loss at Peak Load | Required Shunt Compensation | | |
|---|---|---|---|---|---|---|---|
| | | | | | Peak Load | Dominant Load | Light Load |
| I | 2 | 2 | 1115 MW | 420.8 MW | 1100 Mvar Capacitor<br>550 Mvar Reactor | 2700 Mvar Reactor | 7200 Mvar Reactor |
| II | 3 | 1 | 865 MW | 390.5 MW | 1000 Mvar Capacitor<br>750 Mvar Reactor | 3200 Mvar Reactor | 7600 Mvar Reactor |
| III | 1 | 3 | 520 MW | 359.0 MW | 1000 Mvar Capacitor<br>1100 Mvar Reactor | 3550 Mvar Reactor | 8050 Mvar Reactor |
| IV | 2 | 1 | 620 MW | 366.48 MW | 950 Mvar Capacitor<br>550 Mvar Reactor | 3050 Mvar Reactor | 7400 Mvar Reactor |
| V | 1 | 2 | 400 MW | 348.50 MW | 900 Mvar Capacitor<br>800 Mvar Reactor | 3200 Mvar Reactor | 7700 Mvar Reactor |
| VI | 1 | 1 | 130 MW | 329.12 MW | 1000 Mvar Capacitor<br>550 Mvar Reactor | 3050 Mvar Reactor | 7500 Mvar Reactor |

TABLE XII

TOTAL COST OF EACH TEP CASE

| TEP Case | Cost of Line, Bay, and Reactor | | | Cost of Power Loss | | | Total Cost of TEP $(Million) | Avg. cost per MW power Delivery at bus 18 $(Million) |
|---|---|---|---|---|---|---|---|---|
| | Line Cost $ (Million) | Bay Cost $(Million) | Additional Reactor cost $(Million) | Capital Investment $(Million) | Fuel Cost $(Million) | O&M cost $(Million) | | |
| I | 3561.134 | 63 | 9.450 | 103.584 | 439.463 | 139.895 | 4316.527 | 3.871 |
| II | 3514.909 | 63 | 18.900 | 72.072 | 305.771 | 97.337 | 4071.989 | 4.708 |
| III | 3607.360 | 63 | 29.531 | 39.312 | 166.784 | 53.092 | 3959.080 | 7.614 |
| IV | 2647.738 | 49 | 14.175 | 47.091 | 199.788 | 63.599 | 3021.391 | 4.873 |
| V | 2693.963 | 49 | 21.262 | 28.392 | 120.455 | 38.345 | 2951.418 | 7.379 |

| VI | 1780.567 | 35 | 16.537 | 8.2368 | 34.945 | 11.124 | 1886.411 | 14.511 |
|---|---|---|---|---|---|---|---|---|

Since the system after TEP must operate reliably under all single contingencies, the results indicate that having fewer line connections to supply new loads increases the per MW power transfer cost. Additionally, increasing the number of line connections from one particular bus, rather than making a proportional number of connections from the two nearest possible buses, also raises the per-MW power transfer cost to a new location.

Therefore, the results highlight that to efficiently supply new loads at new locations and maintain system reliability under contingencies, establishing a proportional number of line connections from the nearest buses is the most feasible option.

A preliminary version of this work was presented as a conference paper [39]. We extended that work in this paper, including 1) using accurate line parameters as discussed in Section 2.3, and 2) adding case studies for TEP in Section 5. We also presented another test system in [40], which is distinct from the one introduced here. The test system in [40] incorporates two voltage levels (500 kV and 765 kV), whereas this paper focuses solely on a 500 kV system. As a result, the two test systems differ not only in voltage levels but also in the number of buses and transmission lines. These structural variations lead to different bus load configurations, as shown in the tables of both papers, and yield entirely different load flow outcomes. In addition, the test system in [40] was designed exclusively for peak load conditions and is not applicable to dominant or light loading scenarios. By contrast, the test system proposed in this paper is more comprehensive, supporting analysis under peak, dominant, and light loading conditions. Finally, the TEP studies conducted in the two papers are entirely different, further underscoring that the works are distinct and non-overlapping. The test system presented in this paper is the result of a step-by-step development process documented in our previous works [41–43]. This system, or its variants, has been utilized in transmission expansion studies to compare the performance of unconventional HSIL lines with conventional transmission lines [44–49].

## VI. Conclusion

This paper has introduced a 500 kV, 17 bus test system designed specifically, but not limited to, transmission expansion planning (TEP) studies. The prevailing issue within the existing test cases for TEP purposes is their inability to operate under various loading conditions and ensure feasible system operation across all single contingencies. Additionally, existing systems lack clear information on transmission lines, such as line length and line parameters. In addressing this issue, this paper proposed a test system comprising long transmission lines and meticulously modeled in a distributed manner to determine accurate line parameters. The proposed test system demonstrates effective operation under normal and all single contingency conditions across three distinct loading scenarios: peak load, dominant load, and light load, meeting all technical requirements. Validations of the operation of the test system under each condition are achieved through comprehensive load flow analysis using PSS/E and MATLAB coding. All the detailed information about the proposed system is provided in the paper. Moreover, the analysis of numerous TEP cases using the proposed test system showcases its suitability for extensive transmission expansion planning studies and to benchmark the most appropriate approach for practical implementations. This test system would serve as a valuable reference resource for the system planner and researchers involved in TEP endeavors. This test system can also be used in the steady-state analysis of the power system.

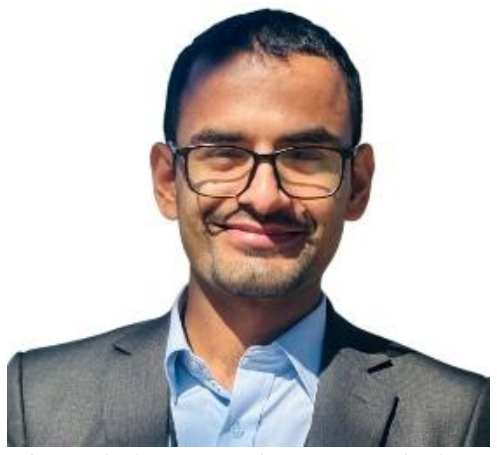

**Bhuban Dhamala** (Graduate Student Member, IEEE) received the B.E. and M.Sc. degrees in electrical engineering from Tribhuvan University, Nepal, and the PhD degree in electrical engineering from The University of Texas at Dallas, USA, in 2025. His research interests include power system modeling and analysis, transmission system planning, power system resiliency, optimization techniques, electricity markets, and the application of AI/ML in power system operations and decision making.

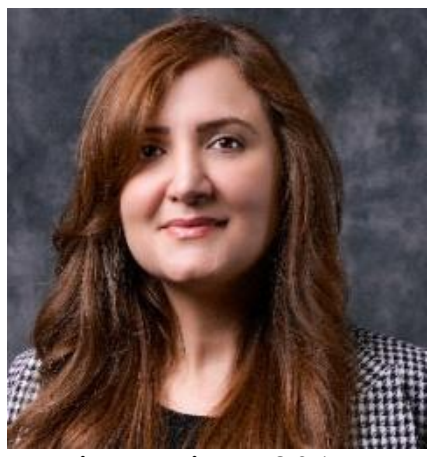

**MONA GHASSEMI** (Senior Member, IEEE) received the Ph.D. degree (Hons.) in electrical engineering from the University of Tehran, Tehran, Iran, in 2012. From 2013 to 2015, she was a Postdoctoral Fellow with the NSERC/ Hydro-Québec/UQAC Industrial Chair on Atmospheric Icing of Power Network Equipment (CIGELE), University of Québec at Chicoutimi (UQAC), Canada. She has been a Registered Professional Engineer since 2015. From 2015 to 2017, she held a postdoctoral position with the University of Connecticut, Storrs, CT, USA. In 2017, she joined the Bradley Department of Electrical and Computer Engineering, Virginia Tech, as an Assistant Professor. In 2021, she was named both the Steven O. Lane Junior Faculty Fellow and the College of Engineering Faculty Fellow at Virginia Tech. In 2022, she joined the Department of Electrical and Computer Engineering, The University of Texas at Dallas (UT Dallas), as an Associate Professor, and was appointed Chairholder of the Texas Instruments Early Career Award. Since 2025, she has been a Professor at UT Dallas. She has authored more than 250 peer-reviewed journal and conference papers, and four book chapters. Her research interests include dielectrics and electrical insulation, high voltage engineering, power systems, and plasma science. She serves as the Vice President −Administrative of the IEEE Dielectrics and Electrical Insulation Society (DEIS); a DEIS Representative on the IEEE-USA Technology Policy Council's Research and Development Policy Committee; an Active Member of several CIGRÉ Working Groups and IEEE Task Forces; and a Technical Committee Member on Dielectrics and Electrical Insulation for Transportation Electrification. She also serves on the Education Committees of both IEEE DEIS and the IEEE Power and Energy Society (PES). Her previous service includes roles as Vice President−Technical of IEEE DEIS, a member of the Nominations and Appointments Committee and an At-Large Member of the Administrative Committee of IEEE DEIS, and a DEIS Representative to the IEEE-USA Committee on Transportation and Aerospace Policy (CTAP). She was a recipient of three of the most competitive, prestigious early-career awards in the United States: U.S. Department of Energy (DOE) Early Career Research Program Award, the National Science Foundation (NSF) CAREER Award, and the Air Force Office of Scientific Research (AFOSR) Young Investigator Research Program (YIP) Award. She has also received a Contribution Award from IET High Voltage and four Best Paper Awards. She serves as an Associate Editor for IEEE Transactions on Dielectrics and Electrical Insulation, IEEE Transactions on Industry Applications, IET High Voltage, International Journal of Electrical Engineering Education, and Power Electronic Devices and Components. She is also a Guest Editor for Aerospace and Energies, and an Associate Guest Editor for the IEEE Journal of Emerging and Selected Topics in Power Electronics and IEEE Transactions on Power Electronics.